# Discovery and rectification of an error in high resistance traceability at NPL: a case study in how metrology works

Speaker: Stephen P. Giblin,

National Physical Laboratory, Hampton Road, Teddington, Middx. TW110LW, U.K.

Phone: +44 20 8943 7161.

Email: stephen.giblin@npl.co.uk.

Authors: Nick E. Fletcher and Colin H. Porter

National Physical Laboratory, Hampton Road, Teddington, Middx. TW110LW, U.K.

**Abstract:**

We broach a seldom-discussed topic in precision metrology; how subtle errors in calibration processes are discovered and remedied. We examine a case study at the National Physical Laboratory (NPL), UK, involving the calibration of DC standard resistors of value 100 MΩ and 1 GΩ. Due to an oversight in the assessment of error sources in the cryogenic current comparator (CCC) ratio bridge used for resistance calibrations, results from the period 2001 to 2015 were in error by approximately 0.7 parts per million (ppm), with quoted uncertainties ($k$=2) of 0.4 ppm and 1.6 ppm respectively. International inter-comparisons did not detect the error, mainly because the uncertainty due to the transportation drift of the comparison standards was too large to resolve it. Likewise, research into single-electron current standards relied on traceability to 1 GΩ, and did not detect the error because at this resistance value it was on the borderline of statistical significance. The key event was a comparison between PTB (Germany) and NPL (UK) of a new small-current measuring instrument, the ultrastable low-noise current amplifier (ULCA). The NPL measurements took place over one week in early 2015 and involved calibrating the transresistance gain (nominally $10^9$ V/A) of the ULCA. At that time, the transport stability of the ULCA was not well established. Nevertheless, calibrations of the ULCA at NPL using a 100 MΩ resistor were sufficiently discrepant with the PTB calibrations to motivate a thorough investigation into the NPL traceability chain, which uncovered the error. All recipients of erroneous calibration certificates were notified, but their responses indicated that the size of the error did not impact their business significantly. This instructive episode illustrates a positive interplay between calibration and research activities and shows that cutting-edge calibration uncertainties must be supported by a vigorous research programme. It is also important for NMIs to maintain a comfortable buffer (at least a factor of 10) between their claimed uncertainty and the uncertainty that their customers require, so that small errors can be resolved without significant impact on measurement stakeholders.

## 1. Introduction

How are measurement errors found and remedied? This is a pertinent question for practitioners of metrology at all levels. The process of accreditation should ensure that measurement services are performed according to technically sound procedures, and at the level of National Metrology Institutes (NMIs), intercomparisons demonstrate equivalence between laboratories. These two processes, accreditation and intercomparison, are the bedrock on which we build trust in our measurements. These systems are not infallible though. An audit by an accrediting body may not be comprehensive enough to pick up subtle errors in a measurement procedure. Indeed, an auditor is unlikely to have time to check every technical detail of a procedure and is not expected to do so. Likewise, intercomparisons may be limited in their effectiveness by the stability of the transport standards, the level of rigour with which measurement protocols are defined and the long time they can take to complete. If we suppose an undetected error exists in a measurement process, we can ask the question: what kind of discrepancy would prompt the metrologist to take extra-ordinary action to investigate the apparatus? In other words, what type of data would the metrologist need to see in order to stop and think, something is not right here.

In this paper we address some of these questions by describing a case study in which an error in a routine calibration service went undetected for 14 years. The error was eventually found, not through an audit or internal procedure review, nor from participation in a formal inter-comparison, but through research into new, more stable types of measuring instrument. Our narrative bears on another more general question relevant to the organisation of metrology labs: to what extent should routine calibration activity and research be combined in the same laboratory? On one level, these activities seem to demand different mind-sets. Routine calibration is mostly the disciplined repetition of a carefully documented procedure, while research involves trying new things and asking questions. However, our story shows how mixing the two activities in the same laboratory, and sharing apparatus, can be mutually beneficial and a catalyst for advancing the state-of-the-art.

Case studies detailing measurement errors are rare but extremely useful. The authors are aware of a well-documented example in the field of ionising radiation measurement which led to a better understanding of mechanical aspects of radioactivity measurements [1]. In the more rarefied world of fundamental constants, attempts to measure the Newtonian gravitational constant, "big G", provide an ongoing case study of discrepant results which show that errors can be present even in carefully designed experiments [2]. Our case study, in contrast, concerns electrical measurements, specifically electrical resistance which is a property of materials describing how easy or difficult it is to pass electric current through them. This is not intended to be a technical paper, and we do not presuppose much background knowledge beyond Ohm's law (which defines resistance as the voltage across a conductor divided by the current through it). Readers with less technical background could skip section 3 without losing the key points of the paper.

## 2. Measurement of electrical resistance

For many decades, calibration of electrical resistance standards with direct current (d.c.) has been a well-established staple service of NMIs. These standards cover an astonishing range of resistance values, from roughly $10^{-5}\,\Omega$ to $10^{15}\,\Omega$. The lower end of the scale is used mainly for measuring large currents in the power generation industry, while the high end finds applications in characterising electrical insulators and semiconductors. Calibrations in the middle range fan out to numerous applications in electrical engineering, manufacturing and research. Since 1990, the resistance scale at many NMIs, including the National Physical Laboratory (NPL), UK, has been formally anchored to an absolute reference resistance based on the quantum Hall effect [3]. This effect generates a resistance of approximately 12.9 k$\Omega$ and scaling up and down by many orders of magnitude to cover the required range of calibration values is the main technical challenge facing a laboratory. Resistance standards (like standards in many other metrology areas) are designed to be nominally equal to decade values – 1 $\Omega$, 10 $\Omega$, 100 $\Omega$ etc – and scaling, using Hamon ratio networks, voltage ratio bridges, room temperature current comparators, or cryogenic current comparators (CCCs) is usually done with sequences of 1:10 or 1:100 ratio measurements.

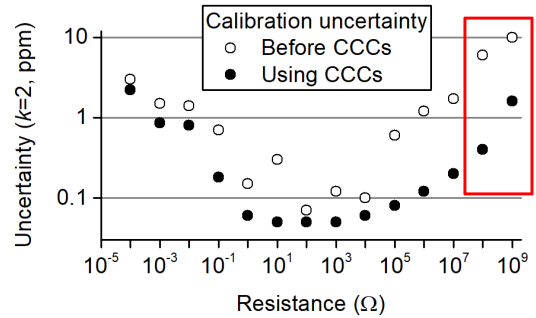

Figure 1. Relative expanded ($k$=2) uncertainty in parts per million as a function of resistance value, for d.c. resistance calibrations at NPL. Open circles: pre-CCC uncertainties. Filled circles: after adoption of CCCs. All uncertainties are published on the calibration and measurement capability (CMC) database.

The most accurate resistance scaling measurements are performed using cryogenic current comparators. These instruments exploit the principle of superconductivity to establish an extremely accurate ratio of electric currents, from which the resistance ratio can be calculated. The principle of the CCC was first demonstrated in the early 1970s [4], and in the early 1990s NPL deployed the first CCC system for routine resistance calibrations [5]. By the early 2000s and following an intensive R&D programme, NPL was performing all resistance calibrations using three different CCC systems, one each specialised for low, medium, and high resistances. The uncertainties offered by the new CCC-based calibrations are plotted in figure 1, with the earlier, pre-CCC, uncertainties ($k$=2) for comparison. It can be immediately seen that CCCs provided a significant reduction in the calibration uncertainty offered to customers across a very wide range of resistance values. The reduction was most dramatic for higher resistances, and we can now focus on the specific area of our case study; standard resistors of nominal value 100 M$\Omega$ and 1 G$\Omega$. These measurements are made on the NPL "high resistance" CCC bridge [6], which was used, starting in 2002, to calibrate resistors from 100 k$\Omega$ up to 1 G$\Omega$.

## 3. Description of the error

Our aim is not to give an in-depth technical analysis of the error, nor is the reader expected to understand the detailed operation of CCC bridges. Fortunately for the purposes of explanation, the NPL high resistance CCC is operated in a very simple mode for calibration of 100 M$\Omega$ and 1 G$\Omega$ resistors, illustrated in figure 2 and described in detail in Ref. [7]. Both resistance values are measured against a 10 MΩ standard, with the bridge configured for a 1:10 (for a 100 MΩ calibration) or 1:100 (for a 1 GΩ calibration) ratio. The current from a single voltage source divides in the two arms of the bridge, and the superconducting comparator measures the ratio of the two currents $I_1/I_2$. In figure 2, $R_1$ is the 10 MΩ reference resistor and $R_2$ is the unknown resistor of either 100 MΩ or 1 GΩ nominal value. Both $R_1$ and $R_2$ are two-terminal standards, meaning they have just two wires connecting the resistance element to the measuring circuit. The complication is that the standard resistors are in series with parasitic circuit resistances $r_1$ and $r_2$, consisting of connecting leads and wiring, and the CCC current ratio $I_1/I_2$ only contains information about the total resistance in the bridge arms: $I_2/I_1 = (R_1+r_1)/R_2+r_2)$.

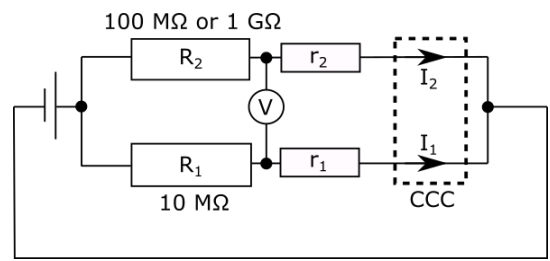

Figure 2. Simplified circuit diagram of the CCC bridge used to measure a 100 MΩ or 1 GΩ standard $R_2$ in terms of a known 10 MΩ standard $R_1$. The additional series resistances due to connecting leads and other wiring are denoted $r_1$ and $r_2$.

To extract the ratio of the standard resistors $R_2/R_1$, the bridge circuit includes a voltmeter. The voltmeter signal combined with the CCC current ratio yields the required ratio of the standard resistors while eliminating the effect of the parasitic resistances [7]. The error arose because the voltmeter data was not used correctly at the data analysis stage to compute the value of $R_2$ from the raw bridge data. To a good approximation, the calculated value of $R_2$ was too low by a relative amount $r_1/R_1$. The error was specific to the resistance values 100 MΩ and 1 GΩ. For the lower values of resistance measured on this bridge, servo control of the CCC current ratio was used resulting in a different bridge equation, and in this case the analysis software was correct. Fortunately, the voltmeter signal was saved in all raw calibration data files, and the data files were saved and archived without any processing. This allowed the data files to be re-analysed, and correct values of $R_2$ to be calculated after the error was discovered.

The parasitic resistances $r_1$ and $r_2$ include the leads connecting the standard resistors to the CCC electronics and the electronics to the superconducting comparator. They also include wires and relay contacts internal to the electronics. The presence of relay contacts is important to the specifics of our problem, because relays may not have the same resistance every time they are opened and closed again. The value of $r_1$ was usually in the range 5 Ω to 10 Ω, yielding a relative error in the value of $R_2$ between 0.5 parts per million (ppm) and 1 ppm. This can be compared to the expanded ($k=2$) uncertainties already stated, of 0.4 ppm for 100 MΩ and 1.6 ppm for 1 GΩ. Clearly, for the case of 100 MΩ the error was larger than the claimed uncertainty.

4. Performance in an intercomparison

To summarise the previous technical section, starting in 2002 calibrations of 100 MΩ and 1 GΩ standard resistors at NPL were in error by between 0.5 and 1 ppm, with quoted relative uncertainties ($k=2$) of 0.4 and 1.6 ppm respectively. If the error had been much larger, of order tens of ppm, it might have manifested in measurements of well-characterised resistors as a discrepancy with measurements made on the old measurement system, which offered uncertainties of 6 ppm and 10 ppm respectively. However, the error was too small to be noticed in this way and calibrations of customers' standards as well as NPL's internally maintained standards continued.

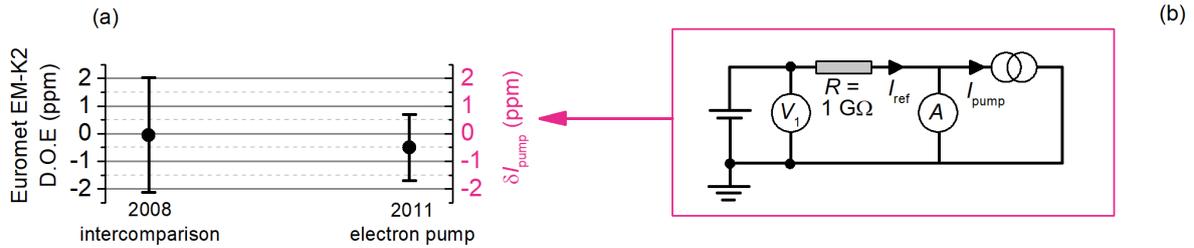

Figure 3. (a): Two data points which did *not* lead to discovery of the CCC error. Error bars are standard (*k*=1) uncertainties. The left data point (left axis) is the degree of equivalence for NPL's participation in the Euromet EM-K2 comparison of 1 GΩ resistance standards [8]. The right point (right axis) is the electron pump current measured using the circuit in (b) and reported in Ref. [10]. (b): Simplified circuit diagram showing the measurement of an electron pump current $I_{pump}$ in terms of a reference current $I_{ref}$.

In 2006, NPL participated in Euromet comparison EM-K2 [8]. Two resistance values were circulated among the 21 participants; 10 MΩ and 1 GΩ. As already noted, the error did not affect the 10 MΩ results as the CCC operated in a different mode for this resistance value. Would the error show up in the 1 GΩ measurements? In other words, would the comparison fulfil one of its intended functions in exposing measurement errors in a participating NMI? The size of the error was roughly the same as the *k*=2 uncertainty (0.8 ppm), and a discrepancy at this level should be enough to prompt a careful review of the measurement system.

The left data point in Fig. 3(a) shows the NPL degree of equivalence (DOE) result for the comparison at 1 GΩ, with the error bar showing the standard (*k*=1) uncertainty. Statistical analysis of inter-comparisons is a complex and evolving subject, but the DOE is an accepted indicator of how well a laboratory's result agrees with the other laboratories. The uncertainty in the DOE reflects the uncertainty NPL reported for its measurement of the comparison standard, but also the uncertainty of the other laboratories as well as additional uncertainty associated with the drift of the transport standards during the timeframe of the comparison. Clearly NPL's DOE does not indicate any discrepancy within the expanded *k*=2 uncertainty of 4 ppm.

The uncertainty in the degree of equivalence is 2.5 times larger than NPL's Calibration and Measurement Capability (CMC) uncertainty declaration. The reason for this is mostly a much larger than usual scatter in the repeated measurements of the transfer standard made at NPL over 2 weeks. Correcting for the temperature coefficient of the standard (which was larger than typical for 1 GΩ standard resistors) also increased the uncertainty, so that the combined *k*=2 uncertainty submitted for NPL's results in the comparison was 3.98 ppm. The five lowest uncertainties (*k*=2) submitted by other participants were 3.8, 4.6, 4.7, 6.1 and 6.7 ppm. Most of the participants reported uncertainties of order 10 ppm or larger. The uncertainty in the DOE cannot be smaller than the smallest uncertainties reported by other participants. Therefore, even if the repeatability problem had not occurred in the NPL measurements and NPL had been able to report an uncertainty close to 1.6 ppm, the uncertainty in the DOE would still have been close to 4 ppm, and the comparison would not have uncovered the error in the NPL measurements.

As a coda to this section, more recent research at NPL has investigated the stability of standard resistors of 100 MΩ and 1 GΩ value on timescales of days to weeks [9]. As discussed further in section 6, We found that the commonly used types of commercially manufactured standards exhibit short-term fluctuations of order 0.5 ppm. This finding sets limits on how well these standards can perform as transfer devices and might partly explain the scatter in the NPL measurements of the comparison resistors 13 years ago. It also raises the possibility that even if the uncertainties of the other participants had been lower, a comparison using these types of transfer standard may not have uncovered an error of order 1 ppm. NPL participated in the next relevant intercomparison, K2.2012, making the measurements of the transfer standards in 2015. By this time, the error had been discovered, but it is worth noting that the expanded ($k=2$) uncertainty in the degree of equivalence for NPL was 1.9 ppm. This is a factor of 2 improvement over the Euromet comparison, but probably still not enough to convincingly resolve a discrepancy of order 1 ppm. As events transpired, the discovery of the CCC error would not come through the organisational structure of comparisons, but through involvement of the CCC bridge in research projects as described in the next sections.

## 5. Research into single-electron current standards

Since the late 1990s, NPL has been active in research into single electron current standards, known as electron pumps. These devices aim to generate a precise reference current by transporting electrons one at a time in response to a periodic clock signal. Ideally, they realise the equation $I = ef$, where $I$ is current, $e$ is the electron charge and $f$ is the frequency of the clock signal. The pumps are research devices, and the object of most of the experiments has been to test them by measuring the pumped current as accurately as possible to see if it is really equal to $ef$, or if there are some errors. The dependence of the pump current on various tuning parameters (such as the gate voltages applied to the semiconductor pumps) was, and is, also a key experimental goal. Measuring the small pump current at the ppm level or better poses a metrological challenge; the maximum frequency $f$ is around 1 GHz, producing a current of 160 pA.

Starting in 2010, NPL began combining a 1 GΩ standard resistor with a voltage source and precision voltmeter to make a reference source of small current using the simple relation $I = V/R$. The standard uncertainty in the reference current was dominated by the 0.8 ppm uncertainty in the 1 GΩ resistance, with a smaller contribution (around 0.5 ppm) coming from the calibration of the voltmeter. The unknown electron pump current could then be compared with the reference current using the circuit shown in Fig. 3(b), where the ammeter is measuring a small difference between the pump and reference currents.

The first measurement campaign using this system was published in 2012 [10], demonstrating that the normalised difference of the electron pump current from $ef$, with standard ($k=1$) uncertainty, was (-0.51 ± 1.2) ppm. This data point is plotted as the right-hand point in Fig. 3(a). This was a breakthrough result, the first experimental demonstration of ppm-accuracy in an electron pump, for a continuously generated current. Higher accuracy had been demonstrated by a team at NIST many years previously, for much lower currents, in cyclic charging of a capacitor [11], and single electron counting [12] experiments, but the 2012 NPL result was the first indication that electron pumps had the potential to work as practical current standards.

The electron pump result was, of course, affected by the error in the calibration of the 1 GΩ resistor. Subsequent re-analysis of the calibration data for the resistor showed the error to be 0.7 ppm, so the corrected electron pump result becomes (-1.21 ± 1.2) ppm. The corrected result is one standard deviation away from *ef*. This was not considered to be a statistically significant discrepancy affecting the main conclusions of the study, and the authors of Ref. [10] chose not to submit an erratum. One more electron pump study [13], with a slightly larger combined uncertainty of 1.37 ppm ($k$=1), was published before the 1 GΩ calibration error was discovered, with a similar conclusion after the data was corrected for the error.

The reader might well object to the inclusion of this section in the paper at all. If the object of the experiments was to study the electron pumps as "unknown" devices, then any deviation of the pump current from *ef* should be interpreted as a property of the pump. Surely any other interpretation would be a basic failure of scientific logic! This is to under-estimate the role of doubt in the mind of the scientist, which triggers an ongoing process of questioning what is supposedly known. It is therefore interesting to speculate as to whether the error might have been unmasked through ongoing single-electron research. Data on different designs of pump operated with different tuning parameters, all revealing a borderline 1 ppm discrepancy, would have triggered a process of questioning the reference current. This would have required a large data set in order to robustly resolve sub-ppm discrepancies between the pump current and *ef*. It would also have required a change in perspective on the part of the experimenters, a growing sense that the electron pump was a more accurate and stable standard than the 1 GΩ resistor.

Interestingly, an earlier episode in NPL's electron pump research gives an example of precisely this kind of perspective shift. Before adopting a reference current source based on a 1 GΩ resistor, the pump current was compared with a reference current generated using a capacitor ramp technique [14]. Initial measurements showed the pump generating flat current plateaus which were offset from *ef* by about 50 ppm. The idea that the pump current could be robustly quantised at the wrong value implied that error processes in the pump were independent of the tuning voltages, and this was felt to be an implausible scenario. The experimental team started to question the accuracy of the reference current generator and discovered a hitherto-unknown frequency dependence in the reference capacitor that, once measured, accounted for the 50 ppm error [15]. The electron pump, initially considered the device under test, had switched roles, revealing imperfections in the apparatus previously assigned the role of reference. Thus, it is possible that an electron pump measurement campaign could have focussed attention on a possible error in the resistance traceability, leading to discovery of the error. However, this didn't happen because the error was discovered before the accumulation of a large enough electron pump data set.

## 6. Discovery of the error

By early 2015, the high resistance CCC, with its attendant error, was at the intersection of a vigorous programme of research activity in addition to performing regular calibrations. It was supporting the single-electron research programme and also investigating the stability of standard resistors. The CCC was operated by both research and calibration staff, and running measurements almost continuously, 24 hours a day, with routine calibrations typically run during the daytime and measurements supporting research overnight. Into this busy schedule came another set of measurements, to test a new type of sensitive ammeter.

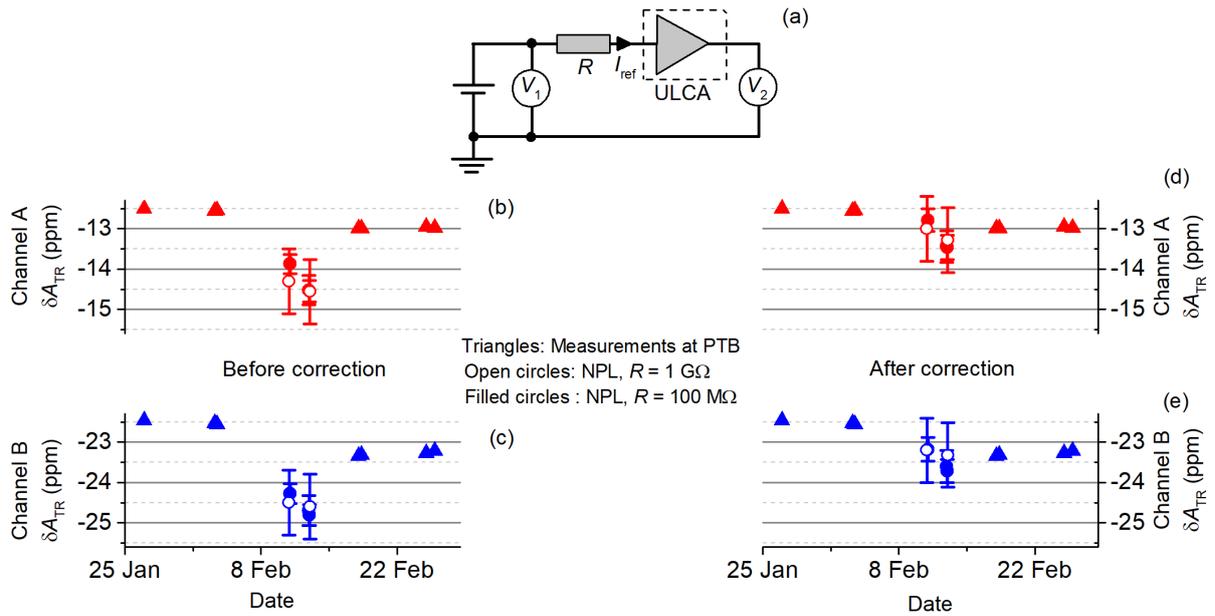

Figure 4. (a): Simplified circuit diagram showing the calibration of the ULCA current-to-voltage gain at NPL, reported in Ref. [17]. (b) and (c): values of the ULCA gain measured at PTB (triangles) and NPL (circles) in early 2015. The NPL data is shown as first analysed, including the error in the resistance traceability. The error bars show the standard ($k$=1) uncertainty and are roughly the same size as the data points for the PTB data. The data is expressed as the deviation of the gain in parts per million from its nominal value of 1 GΩ. (d) and (e): the same data after correcting the error in the NPL resistance traceability. This is a subset of the data that appears as the bottom panel of figure 5 in Ref. [17]

The challenge of measuring small currents, for single-electron research as well as practical calibration applications, had been exercising the minds of researchers at several NMIs. In early 2015, a research team at the Physikalish-Technische Bundesanstalt (PTB) presented a new tool for small-current metrology, the ultrastable low-noise current amplifier (ULCA) [16]. This is a versatile instrument which can be configured as a current source or current meter, but for the purposes of our story, it is the "meter" function that is important. In this mode it acts as a trans-resistance amplifier, converting an input current $I_{IN}$ to an output voltage $V_{OUT}$ according to $V_{OUT} = I_{IN}A_{TR}$, where $A_{TR}$ is the trans-resistance gain of the ULCA. One of the key design features of the ULCA is that $A_{TR}$ should be very stable. After calibrating, it should not drift by more than 1-2 ppm in a year, and it should not be affected by environmental factors or transportation. To put this feature into perspective, it represents roughly a factor 100 improvement over the stability of commercial trans-resistance amplifiers. Of course, the stability under transportation had to be tested experimentally, and several NMIs agreed to calibrate ULCAs transported from PTB. Calibrations at PTB before and after shipment to another NMI would reveal any shifts in $A_{TR}$ due to transport effects. Logistically, this looks very much like an intercomparison, but the goal was different: with the ULCA transport, it was the behaviour of the transfer standard which was under investigation, not the calibration capabilities of the participating NMIs whose measurements were assumed to be correct within the stated uncertainties.

NPL first calibrated the gain of a 2-channel ULCA (essentially two separate ULCAs in one box) in early February 2015. For these calibrations, the same reference current source from the single-electron research programme was used to feed a current into the ULCA input. Referring to the circuit diagram in figure 4 (a), $I_{ref} = V_1 / R$, and $A_{TR} = V_2/I_{ref}$, so $A_{TR} = RV_2/V_1$. The voltmeters measuring $V_1$ and $V_2$ were calibrated frequently against a primary standard with uncertainties of a few parts in $10^7$. Both 100 MΩ and 1 GΩ resistors were used for $R$.

The $A_{TR}$ results calculated at the time of the NPL measurements for the two ULCA channels are shown in figures 4 (b) and (c). The markedly lower uncertainty for the PTB calibrations was possible by calibrating the ULCA gain in a 2-stage process directly using a CCC [16], a technique which was only developed at NPL several years later. After the ULCA had been returned to PTB and calibrated a few times there, it was clear that the NPL measurements were about 1.5 ppm below the PTB measurements for both ULCA channels (the two channels are electrically independent, but mechanically share the same case). This was one of the first ULCA transportation tests, and the transport stability of the ULCA was not at all well established at this time. Nevertheless, it seemed implausible that *both* channels should experience the same jump *down* in gain on the first transport to NPL, and then a jump back *up* on the second transport back to PTB. An error in the NPL reference current was the simplest explanation for the data, and it would have to be an error which affected both 100 MΩ and 1 GΩ resistors by an almost equal amount. This data was the smoking gun that triggered an investigation of the CCC bridge, and the error was found within a few days of starting the investigation. The ULCA calibration data was re-analysed using the corrected values of the 1 GΩ and 100 MΩ resistors (figure 4 (d) and (e)), and subsequently published along with data from LNE (France) with the conclusion that the ULCA gain was robust at the 1 ppm level during international transportation [17]. A few years later, another NPL – PTB study on ULCA transportation improved this number to 0.2 ppm [18], and other studies on the ULCA have shown it to have sub-ppm stability on annual time-scales and under a range of transportation conditions [19].

At roughly the same time as the ULCA measurements in early 2015, NPL scientists were starting to investigate short-term fluctuations in the resistance of 100 MΩ and 1 GΩ resistance standards, as it was becoming clear that these fluctuations were limiting the uncertainty achievable in the electron pump measurements. On timescales of hours to a few days, commercial standard resistors which use thick-film elements show fluctuations as large as 1 ppm with a roughly $1/f$ power spectrum. In plainer language, the resistance drifts around, even if the resistor is very well temperature controlled. It was concluded eventually that this drift behaviour is a fundamental property of conduction in thick-film elements [9], but when it was first noticed this conclusion was not obvious and the CCC bridge was suspected to be at fault. A programme of measurements, mostly of 100 MΩ resistors, was being undertaken with the CCC in various simplified configurations, eliminating parts of the circuit path used in the "standard" calibration configuration. It is highly likely that if the error had not been uncovered by the ULCA comparison, it would have been discovered by these test measurements.

## 7. Customer contact

Immediately following discovery of the error, all calibrations performed on the bridge since 2002 were re-analysed, and the correct value of the resistors under test calculated. Fortunately, all calibrations at NPL were, and are, performed under the auspices of an ISO 9001 quality system. All raw data is filed electronically in a standardised format and calibration records make it easy to link a calibration certificate to the raw data. Another important factor is the clear separation between the data acquisition and analysis stages. Since the error was in the analysis stage, the analysis software could be updated and re-run on the same raw measurement data. A total of 93 calibrations were affected, for 17 distinct customers including 3 NMIs. Letters were sent to all the customers with details of the affected calibration certificates, and the size of the error. Figure 5 shows a histogram of the errors reported to customers for resistors calibrated at the usual test voltage of 100 V (calibrations at lower test voltages are not included in this data). The distribution of errors reflects the distribution of relay contact resistances in the circuit path. Most of the errors clustered around the 1 ppm level, with a small number between 2 ppm and 2.5 ppm.

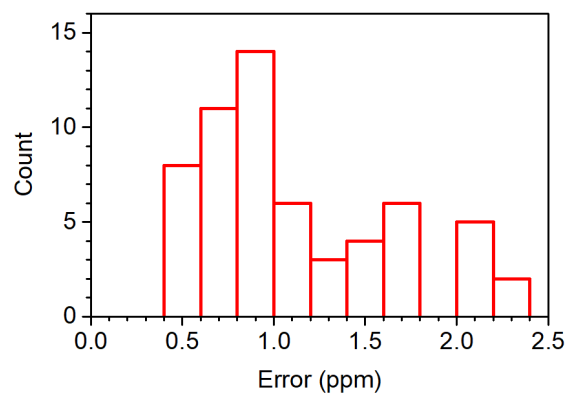

Figure 5. Histogram of the error in 100 MΩ and 1 GΩ resistor calibrations (at a test voltage of 100 V) reported to customers.

Without exception, all the customers were satisfied that the correction was sufficiently small not to be a problem. Most of the non-NMI customers were accredited calibration companies. Six of the customers claimed uncertainties in their own calibration schedules below 10 ppm ($k=2$) for 100 MΩ measurements: in increasing order, 1.5, 2, 2.1, 4.5, 5 and 9 ppm. Only the three lowest of these are affected by the NPL error. For all the rest the error presumably had no impact on their capability. In addition to the calibration data, several other measurements performed on the bridge were re-analysed. These include the measurements for the Euromet EM-K2 intercomparison, discussed in section 4, the traceability measurements for the single-electron pump research, discussed in section 5, and of course the measurements of the ULCA gain discussed in section 6.

In the process of re-analysing the raw calibration data, it was noticed that occasionally a data set occurred with an anomalously large error, in excess of 10 ppm. These data sets were invariably rejected, usually before the measurement was complete, due to excess noise or a clear discrepancy with the expected value of the test resistor based on its history. The bridge was reset and the measurement re-started. It can now be understood that these 'anomalous' data sets were the result of large series resistance due to accumulation of contaminants on the relay contacts. Resetting the relays restored a good contact. We acknowledge that the anomalous data sets were indicating a potential source of error, and an investigation into their cause could have uncovered the error at any time in the history of the measurements. The implications of this observation will be discussed in the next section.

## 8. Discussion and lessons learned

As the saying goes, all's well that ends well. A subtle error in a measurement was found and corrected, and for the end users of the measurement, the error was too small to cause a problem. Nevertheless, the fact remains that a calibration service at an NMI operated for 14 years with an error, which was not detected through the mechanism of international intercomparison. The inability of the relevant intercomparison to detect the error was discussed in section 4 and was primarily due to unusual behaviour of the transfer standard. Even if the transfer standard had been perfect, exhibiting no transport effects or drift, the lack of other comparison participants with low enough uncertainty would still have masked the error. The error was unmasked eventually by a bilateral comparison of a research device, the ULCA, with a partner (PTB) who were able to calibrate the ULCA with extremely low uncertainty. Despite a lack of knowledge of the transport stability of the ULCA, the bilateral comparison generated a suspicious data set which turned attention onto the NPL measurement system.

Here, we summarise one at a time the important lessons learned from this episode; the factors which enabled us to discover the error, as well as actions which could have led to discovering it earlier.

**Investigate anomalous readings**. It is tempting to attribute highly anomalous measurements rather vaguely to "noise", "interference" or a number of other causes. If the problem goes away when instruments are reset or power cycled, it is even more tempting to get on with the measurements and forget about the anomalous data. As mentioned at the end of the previous section, anomalous measurement in our case contained the key to finding the error, had they been thoroughly followed up.

**Consider a bilateral inter-comparison**. A common scenario, indeed, the one that started off this narrative, is that a laboratory has made an improvement to its capability resulting in a reduction in uncertainty and wants to validate the new uncertainty. In this case, a focussed bi-lateral comparison, on a short timescale and with a partner possessing similar measurement capability, is probably a better tool than participation in a multi-partner intercomparison. In our study, the tight timescale and highly focussed nature of the bi-lateral ULCA comparison were as important as the properties of the ULCA in uncovering our error.

**Maintain an "uncertainty buffer".** If you can make measurements 10 times more accurately than your customers need, this is not wasted effort. It provides a buffer zone in which subtle errors can be rectified without impacting the customers' business.

**Participate in research wherever possible.** Using established measurement systems in non-standard ways, or to measure new types of standard, is a great way to validate the performance of these systems and tease out errors if they are present.

**Be prepared to doubt.** An important element of our narrative has been the change in perspective that can take place, when confidence grows that a new unproven type of standard or measuring system is more stable than an existing standard or measuring system. If anomalous data is produced, the simplest explanation is probably the correct one, even if that explanation implicates an error in an established measurement system which is part of a calibration service.

## 9. Summary

We have related how an error, slightly larger than the claimed uncertainty, was detected in a measurement system after 14 years of operation. The error was not detected by an intercomparison but was picked up through measurements of an unproven research device. We have shown how this story illustrates a positive symbiosis between research and calibration activities taking place in the same laboratory. It also forms a powerful argument for NMIs to operate at uncertainties well below what their customers need. We hope it will prove instructive to metrology practitioners at all levels, in managerial as well as technical roles.

**Acknowledgments**

The authors are supported by the UK department for Business, Energy and Industrial Strategy. The single-electron research reported in section 5 was partly supported by project "Reuniam" of the European Metrology Research Programme (grant no. 217257). The ULCA comparison reported in section 6 was partly supported by the Joint Research Project "Qu-Ampere" (SIB07) within the European Metrology Research Programme (EMRP). The EMRP is jointly funded by the EMRP participating countries within EURAMET and the European Union.